\documentclass[doublecol]{epl2} 

\usepackage{amsmath,amssymb,amsthm}
\usepackage{graphicx}
\usepackage[normalem]{ulem}

\title{Isometric immersions, energy minimization and self-similar buckling in non-Euclidean elastic sheets.}
\shorttitle{Isometric immersions and self-similar buckling}

\author{John Gemmer\inst{1,5} \and Eran Sharon\inst{2,5} \and Toby Shearman\inst{3} \and Shankar C. Venkataramani\inst{3,4,5} }
\shortauthor{J. Gemmer, E. Sharon, T. Shearman and S. C. Venkataramani}

\institute{
\inst{1} Division of Applied Mathematics, Brown University, Providence, RI 02906, USA \\
\inst{2} Racah Institute of Physics, The Hebrew University, Jerusalem, 91904, Israel \\
\inst{3} Program in Applied Mathematics, University of Arizona, Tucson, AZ 85721, USA\\
\inst{4} Mathematics Department, University of Arizona, Tucson, AZ 85721, USA\\
\inst{5} {Kavli Institute of Theoretical Physics, University of California, Santa Barbara, CA 93106, USA} \\
}


\abstract
{
The edges of torn plastic sheets and growing leaves often display hierarchical buckling patterns. We show that this complex morphology (i) emerges even in zero strain configurations{, and}  (ii) is driven by a competition between the two principal curvatures, rather than between bending and stretching. We identify the key role of  branch-point (or ``monkey-saddle") singularities in generating complex wrinkling patterns in isometric immersions, and show how they arise naturally from minimizing the elastic energy. 
}

\begin{document}

\maketitle


%
%

\section{Introduction} 
\begin{figure}[ht]
\includegraphics[width=\hsize]{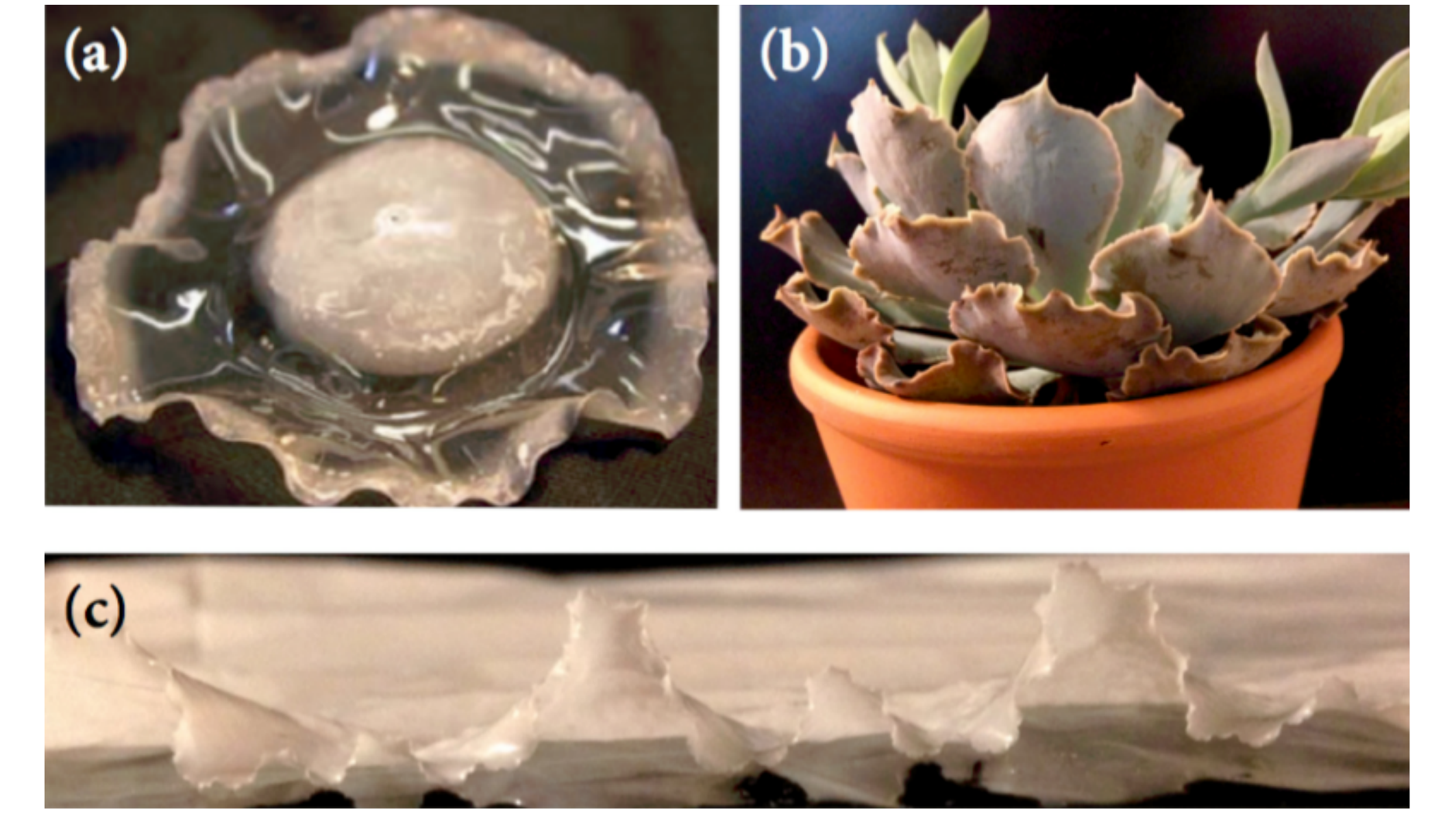}
\caption{Examples of complex wrinkling patterns in swelling thin elastic sheets. (a) Hydrogel disk with non-uniform swelling pattern. (b) Ornamental echeveria house plant. (c) Edge of a torn trash bag. }\label{fig:Examples}
\end{figure}

The rippling patterns observed in torn plastic sheets \cite{sharon2002buckling, audoly2003self, sharon2007geometrically}, leaves \cite{audoly2002ruban, marder2003theory, marder2003shape,liang2009shape} and swelling hydrogels \cite{efrati2007spontaneous, klein2007shaping, kim2012designing} provide striking examples of periodic and self-similar patterns; see fig.~\ref{fig:Examples}. Within the formalism of finite elasticity, such patterns are understood as resulting from the sheet buckling to relieve growth induced residual strains \cite{goriely2005differential, ben2005growth}. On the one hand, complex, self-similar patterns in elastic sheets can arise from boundary conditions that preclude the possibility of relieving in-plane strains \cite{GiOrtz,BCDM, bella2014metric}. On the other hand many growth patterns generate residual in-plane strains which can be entirely relieved by the sheet forming an isometric immersion of a hyperbolic Riemannian metric; indeed  smooth hyperbolic metrics on bounded domains can always be immersed in $\mathbb{R}^3$ by smooth isometries \cite{han-hong-book}  (but not so for unbounded domains \cite{amsler1955surfaces,efimov1964generation,nechaev2001plant}). Why then do we observe self-similar buckling patterns in free elastic sheets? In this letter we report on multiple strands of recent work, that, taken together, address this puzzle. 

We show that a large class of  growth profiles admit smooth ({\em i.e.} infinitely differentiable) as well as (many) non-smooth configurations of the sheet with vanishing in-plane strain. The non-smooth configurations are piecewise surfaces constructed
by gluing together isometries along ``lines of inflection'' and at ``branch points'' in such a manner that the resulting surface does not concentrate bending energy. 
We also 
show that minimizing the bending energy among the various isometric immersions naturally leads to complex, self-similar, wrinkling patterns. 

\section{Non-Euclidean model of elasticity} The non-Euclidean formalism of thin sheet elasticity  \cite{efrati2009elastic}
posits that growth permanently deforms the intrinsic distance between material points, so growth is encoded in a Riemannian metric $\mathbf{g}$.  Specifically, material points on the center surface are labelled by coordinates $(x,y) \in \Omega$, a subset of $\mathbb{R}^2$, 
and the distance between such points is given by the arc-length element:
\begin{equation}
ds^2=g_{11}(x,y)dx^2+2g_{12}(x,y)dxdy+g_{22}(x,y)dy^2.
\end{equation}
By the Kirchhoff hypothesis \cite{solid-mech-book}, the conformation of the sheet as a 3-dimensional object in $\mathbb{R}^3$ is determined by an immersion $F:\Omega \to \mathbb{R}^3$ of the center surface. The equilibrium configuration is then taken to be a global minimizer of an elastic energy modeled as the sum of stretching and bending contributions:
\begin{align}
E[F] &=\mathcal{S}[\gamma]+t^2\mathcal{B}[H,K] \nonumber \\
&=\int_{\Omega} Q(\gamma)\,dxdy+t^2 \int_{\Omega}(4H^2-2K)\,dxdy,
\label{elastic-energy}
\end{align}
where $\gamma = (\nabla F)^T\cdot \nabla F-\mathbf{g}$ denotes the in-plane strains in the center surface, $t$ is the thickness of the sheet, $Q$ is a quadratic form, and $H$ and $K$  are respectively the mean and Gaussian curvatures of the center surface \cite{efrati2009elastic, lewicka2011foppl}. {Provided they exist}, the $t \to 0$ (vanishing thickness) limits of  minimizers of the elastic energy (\ref{elastic-energy}) are necessarily {finite bending energy isometric immersions} \cite{lewicka2011scaling}. 


{For isometries, the Gauss curvature $K = \kappa_1 \kappa_2$, the product of the the {\em principal curvatures} $\kappa_{1,2}$, is determined by the metric. Decreasing $|\kappa_1|$ increases $|\kappa_2|$ and vice-versa so there is a competition between their contributions to the bending content $\int (\kappa_1^2 + \kappa_2^2) dx dy$.  We define a geometric quantity, the {\em disparity} $\eta$, by:
\begin{equation}
\eta \equiv \frac{H}{\sqrt{|K|}} = \frac{1}{2}\left(\sqrt{\left|\frac{\kappa_1}{\kappa_2}\right|} - \sqrt{\left|\frac{\kappa_2}{\kappa_1}\right|}\right) \label{eq:disparity}
\end{equation}
The bending energy density $\kappa_1^2+\kappa_2^2 \sim H^2 = \eta^2 |K|$. Thus $\eta$ quantifies the local contribution to the bending content from the mismatch in the principal curvatures.
}
 
\section{Power law metrics and single wavelength isometries} 
{ We investigate finite bending energy isometric immersions of the strip $\Omega=\mathbb{R} \times [0,W]$ with the metric}
\begin{equation}
\mathbf{g}=(1+2 \epsilon^2 f(y))\,dx^2+dy^2, \quad f(y) = \frac{\alpha}{4 (\alpha+1)} \left(1+\frac{y}{l}\right)^{-\alpha}
\label{eq:StripMetric}
\end{equation}
where  
$\alpha\in (0,\infty)$, $l$ is a length scale and $\epsilon>0$.  These metrics corresponds to $y$ dependent growth in the $x$ direction localized near the $y=0$ edge of the sheet. It includes the metrics considered in \cite{marder2003shape, audoly2003self, sharon2007geometrically, liang2009shape, bella2014metric} as particular cases. 

For $\epsilon \ll1$, approximate isometries are obtained from the F\"oppl - von K\'arm\'an ansatz $F(x,y) = (x+\epsilon^2 u,y+\epsilon^2v,\epsilon w)$. 
In the small slope regime, {\em i.e.} for $\epsilon \ll 1$, the mean and Gaussian curvatures are given by $H=\frac{\epsilon}{2}\Delta w$ and $K=\epsilon^2 \det(D^2 w)$ \cite{audoly2010elasticity}.  If $w$ satisfies
\begin{equation}
\det(D^2 w(x,y))=
-f^{\prime \prime} =  -\frac{\alpha^2}{4}\left(1+\frac{y}{l}\right)^{-\alpha-2}, \label{eq:mng_amp}
\end{equation}
we can solve for 
$u$ and $v$ to 
obtain isometries at $O(\epsilon^2)$ \cite{audoly2010elasticity}.  {It is straightforward to check that eq.~\eqref{eq:mng_amp} has a one parameter family of product solutions 
\begin{equation}
w^0(x,y) = k^{-1} \psi(k x)  \left(1+y/l\right)^{-\alpha/2}
\label{prod_soln}
\end{equation}
where $\psi'^2 + |\psi|^{2\alpha/(2+\alpha)} = 1$. These product solutions for $w$ necessarily yield {\em single wavelength isometries}. The wavelength in the $x$ direction is set by $k$ {\em independent of $y$}, so there is no refinement as $y \to 0$. The product solutions in eq.~\eqref{prod_soln} have finite bending content for $\alpha > 2/3$. Further,}
\begin{equation}
H(x,y)  =\epsilon \left[\frac{c_1(\xi) }{kl^2(1+y/l)^{\alpha/2+2}} - \frac{c_2(\xi) k}{ (1+y/l)^{\alpha/2}}\right],
\end{equation}
where $\xi = k x$, and  $c_{1,2}$ are positive functions, independent of $k$. 
%
Averaging $\epsilon^{-2}H^2$ 
in $x$ we have the $k$ dependence of $\bar{B}$, the normalized (by $\epsilon^2$),  nondimensional, bending content per unit length in the $x$ direction:  
\begin{equation}
\bar{B} \sim C_1 k^2 l \int_0^W \frac{dy}{(1+y/l)^\alpha} +  \frac{C_2}{k^2 l^3} \int_0^W \frac{dy}{(1+y/l)^{\alpha +4}} \label{eq:Bending_Content}
\end{equation}
with $C_1,C_2$ positive constants. Let $k^*$ denote the value of $k$ that minimizes $\bar{B}$;  the optimal ``global" wavelength $\lambda_{glob}$ is given by: 
\begin{equation} 
  {\lambda_{glob} \sim \frac{1}{k^*} \sim l \left|\frac{ \left(1+W/l\right)^{1-\alpha}-1}{\left(1+W/l\right)^{-3-\alpha}-1} \right|^{\frac{1}{4}} ,}
\end{equation}
where we are taking $|(1+W/l)^{1-\alpha}-1|=\ln\left(1+W/l\right)$ for $\alpha=1$.
We can also determine the optimal ``local"  wavelength $\lambda_{loc}(y)$ by minimizing {the average of $H^2$ at a given $y$, (eq.~\eqref{eq:Bending_Content} without integrating in $y$) to obtain}
\begin{equation} 
  \lambda_{loc}(y) \sim l (1+y/l) = (y+l).
  \end{equation}
This is related to the buckling wavelength for a thin strip cut out of the sheet \cite{sharon2007geometrically, liang2009shape} at $y$. $\lambda_{loc}(0) \approx \lambda_{glob}$, but for $W \gg l$, there is a range of $y$ where $\lambda_{loc}(y)$ differs substantially from $\lambda_{glob}$, giving an energetic impetus for the sheet to increase the buckling wavelength as one moves away from the edge at $y=0$.

\section{Branch points in isometric immersions} 
There is, however,  {a geometric obstruction to changing the wavelength in $y$. Smooth surfaces with negative Gaussian curvature are locally saddle-shaped. At every point $p$ on such a surface there are two asymptotic directions given by the intersection between the surface and its tangent plane at $p$. An argument, that we will present elsewhere, proves that surfaces that have precisely two asymptotic lines through every point will necessarily have a single wavelength and cannot {\em sub-wrinkle}. We introduce the notion of ``branch point'' singularities which allow for multiple asymptotic directions at isolated points and help circumvent the obstruction to sub-wrinkling in smooth isometries.}

\begin{figure}[h!]
\begin{center}
\includegraphics[width=\hsize]{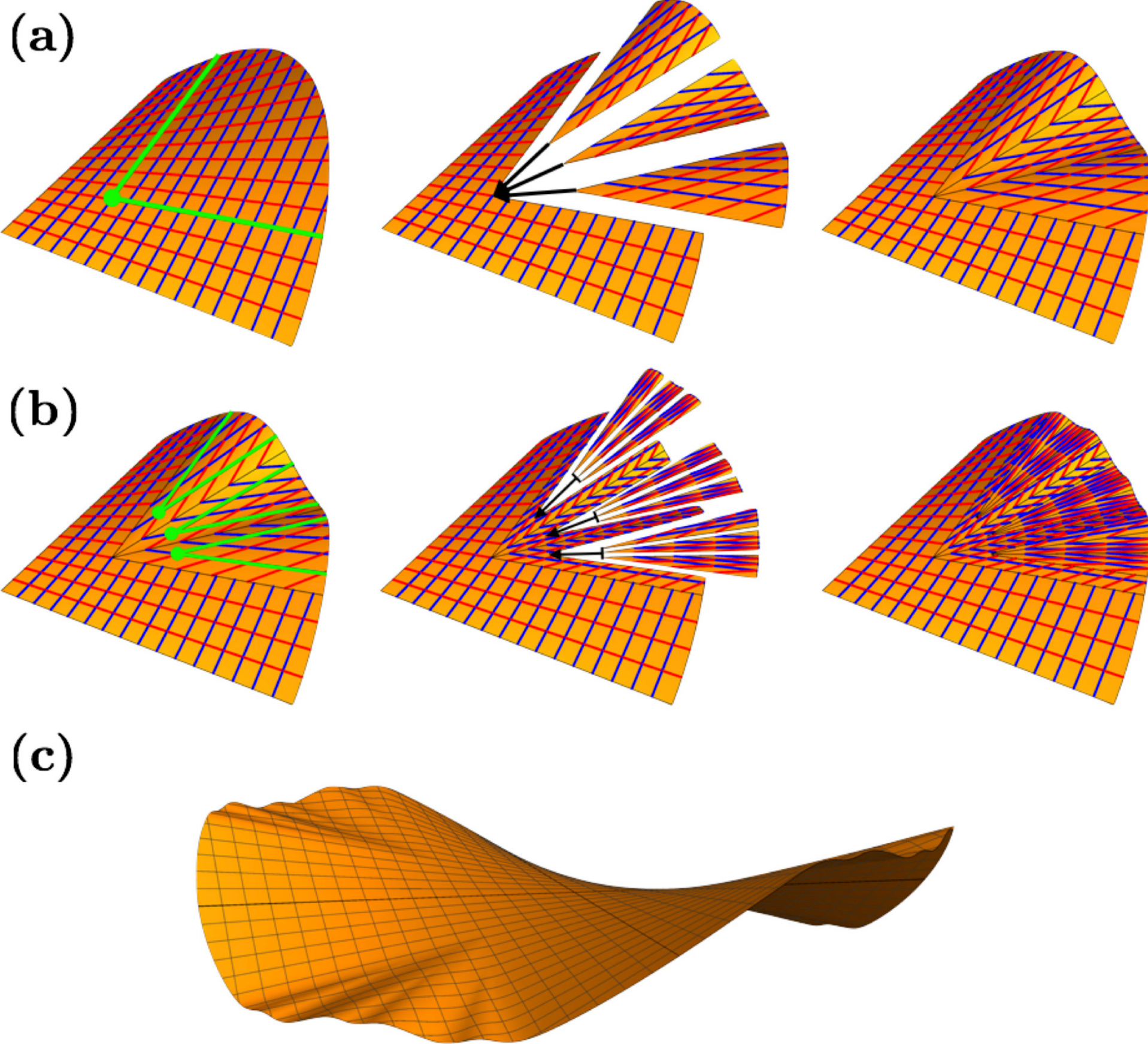}
\caption{Solutions to $\det(D^2 w)=-1$ with branch points. (a) Three sub-wrinkle solution on the first quadrant. (b) Nine sub-wrinkle solution.  (c) Extension to 36 sub-wrinkle solution on the plane.}
\label{fig:SelfSimilar}
\end{center}
\end{figure}

{First, consider solutions of $\det(D^2w) = -1$, the small-slope approximation to the isometry condition for a surface with constant negative Gaussian curvature. Let $(x,y) $ and $(r,\theta)$ denote Cartesian and polar coordinates on $\mathbb{R}^2$. Piecewise solutions to $\det(D^2 w) = -1$ are given by 
$$
w(x,y) = ax + by + c + \begin{cases} y(x-y\cot(\theta_+) ) & 0 \leq \theta \leq  \theta_+  \\ y(x+ y\cot(\theta_-)) & - \theta_- \leq \theta \leq 0 \end{cases} 
$$
where $0 < \theta_{\pm} < \pi$. The surface $z = w(x,y)$ has a continuous tangent plane along the ray $\{\theta= 0\}$ where the two ``pieces" are attached, so bending energy {\em does not} concentrate on this ray. 
This ray is a \emph{line of inflection} -- a principal curvature  is discontinuous across this ray, but the bending energy density is finite.

This construction can be extended to piecewise quadratic surfaces about any point $(x_0,y_0)$. Given $2n$ (an even number) and a sequence of angles $\theta_0 <\theta_1 <\theta_2 < \ldots <\theta_{2n} = 2\pi + \theta_0$, and a plane  $z = a x + by + c$, define lines of inflection on the plane by $x(\lambda) = x_0+ \cos(\theta_i) \lambda$ and $y(\lambda) = y_0 + \sin(\theta_i) \lambda$  with  $\lambda \geq 0$. These lines determine quadratic surfaces with $K = -1$ that can be joined along these lines of inflection. This generalizes our construction in ref.~\cite{gemmer2011shape}; see fig. \ref{fig:SelfSimilar}(a).} 

The {resulting} surface {from this construction} is smooth only if $n = 2, \theta_2 = \theta_0+\pi$ and  $\theta_3 = \theta_1+\pi$. In particular, the surface is not smooth if $n \geq 3$. We {define the point} $(x_0,y_0)$ where the 6 or more lines of inflection meet to be a \emph{branch point} or {\em $n$-saddle} for the surface; $(x_0,y_0)$ is indeed a branch point for the map $(x,y) \mapsto \nabla w$.

Multiple branch points can introduced on the surface recursively as illustrated in fig.~\ref{fig:SelfSimilar}(b-c). The tangent plane at $(x_0,y_0,w(x_0,y_0))$ intersects the surface along 4 rays with azimuthal angles $\varphi_0, \varphi_1, \varphi_2 = \varphi_0+\pi, \varphi_3 = \varphi_1+\pi$. We can introduce a branch point at $(x_0,y_0)$ using the tangent plane as the common plane, and define a {\em daughter sequence} of angles $\theta_0 = \varphi_0 < \theta_1 < \theta_2 < \theta_3 = \varphi_1 < \theta_4 = \varphi_2 < \theta_5 = \varphi_3 < 
\theta_0+ 2 \pi$. {In the sector $\theta_{i-1} \leq \theta \leq \theta_i,$ $H = \eta = \pm \cot(\theta_i-\theta_{i-1})$, where the disparity $\eta$ is defined in eq.~\eqref{eq:disparity}. The sequence $\{0,\frac{3\pi}{4},\pi,\frac{7\pi}{4},2 \pi\}$ and the daughter sequence $\{0,\frac{\pi}{4},\frac{\pi}{2},\frac{3\pi}{4},\pi,\frac{7\pi}{4},2 \pi\}$ are unique in that all the sectors have disparity $\eta = \pm 1$, and thus have the same energy density. These sequences are therefore natural candidates for the local structure at bifurcations in energy minimizers corresponding to adding branch points.} Moreover, this surface displays {\em sub-wrinkling} or {\em refinement} -- the ``number of waves" increases with radius. 

\section{Branch points in the strip geometry} We now return to the strip geometry in eq.~\eqref{eq:StripMetric}; we want to create patterns which have an effective wavelength that increases with $y$, so they have lower energy than the single wavelength isometry. A natural idea is to introduce branch points into the product solutions of eq.~\eqref{eq:mng_amp} to allow for local refinement of the wavelength near $y=0$. However, the product solutions are not ruled by straight lines, and it is not immediately obvious how one adapts the above construction.

We can construct solutions of eq.~\eqref{eq:mng_amp} as follows. Let $Z(X,Y)$ be a (piece of a) surface satisfying $Z_{XX}Z_{YY}-(Z_{XY})^2=-1$ with 
$Z_{XX} \neq 0, Z_{YY} \neq 0$ (true generically by rotating the $XY$-axes if necessary). Define $P = \partial_X Z, Q = \partial_Y Z$, $\sigma = (f''(y))^{1/4}$.  Direct computation shows that the over-determined system \begin{align}
&\partial_Y y   = -\sigma^{-2}, \quad  x   = \sigma^{-1} \left( X + \xi_{+}  + \xi_{-}  \right), \nonumber \\
 & \partial_x w   = P, \quad \partial_y w  =  \sigma \left( Q+ \xi_{+} - \xi_{-}  \right) ,
\label{eq:Backlund}
\end{align}
satisfies $\partial_y(\partial_x w) = \partial_x(\partial_y w), \mathrm{det}(D^2w) = -f''(y)$ if
\begin{equation}
\frac{\partial \xi_{\pm}}{\partial Y} \mp \frac{\partial \xi_{\pm}}{\partial P} = \frac{\partial_Y \sigma}{2 \sigma} \left( X \mp Q + 2 \xi_{\mp} \right),  \mathrm{det}(D^2Z) = -1.
\label{eq:along-characteristics}
\end{equation}
Solving eq.~\eqref{eq:along-characteristics} reduces to integrating ODEs on surfaces satisfying $\mathrm{det}(D^2Z) = -1$. If the underlying surface $Z(X,Y)$ has branch points, then so does the 
surface $w(x,y)$. Consequently there exist many non-smooth solutions to~\eqref{eq:mng_amp} for metrics with $y$-dependent growth. In particular eq.~\eqref{eq:Backlund} yields $(1+y/l) \sim Y^{-2}$ 
for the metric in ref.~\cite{audoly2003self}; in this case, refinement with increasing $Y$ for $Z(X,Y)$ corresponds to sub-wrinking as $y \to 0$. 

\begin{figure}[ht]
\begin{center}
\includegraphics[width=\hsize]{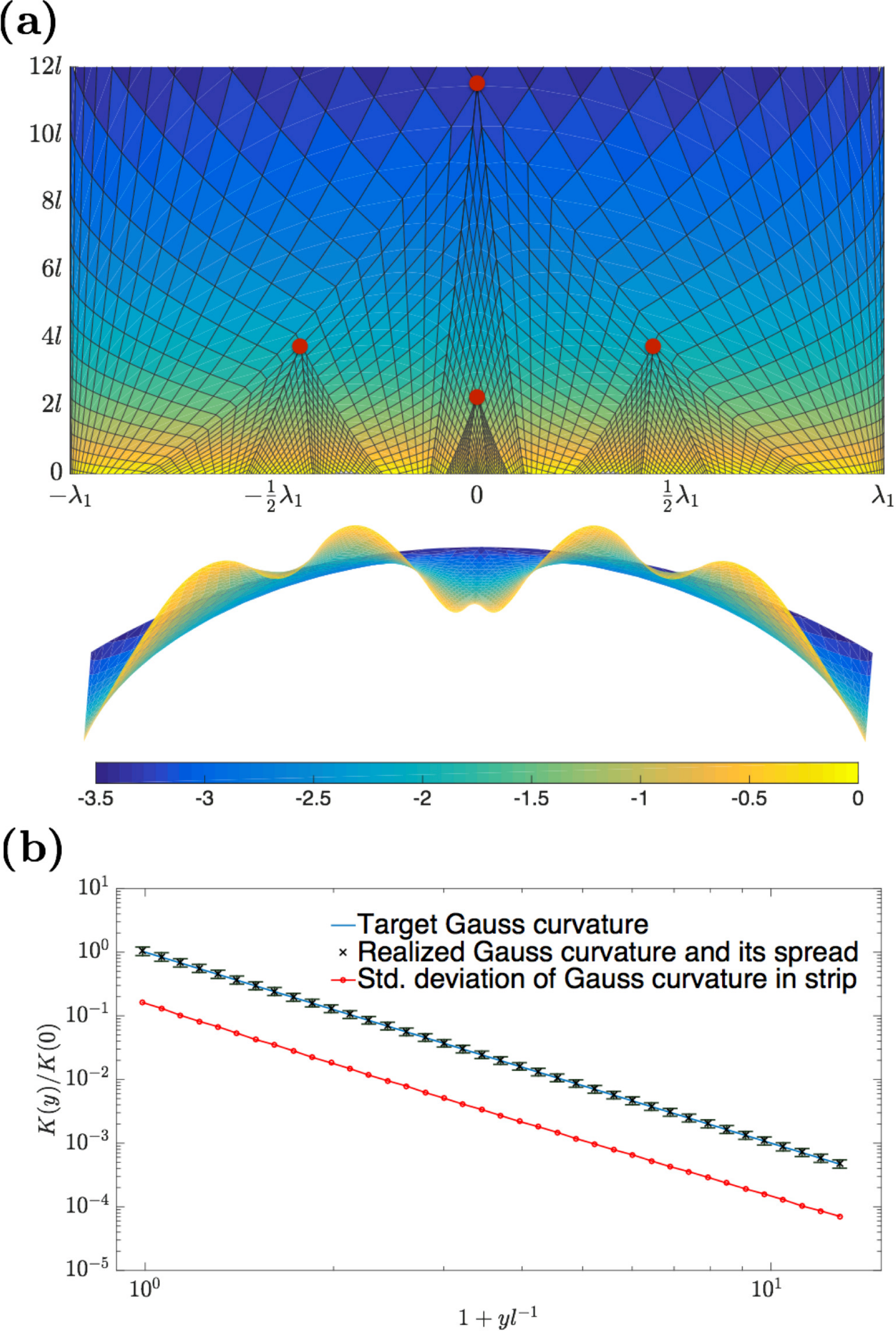}
\caption{(a) Discrete approximation of a four branch point (highlighted)  
isometric immersion for growth profile $f(y) = (1+y/l)^{-1}$. The surface is colored by $\log_{10}(K_d/K(0))$, the logarithm of the discrete curvature; note that the curvature (essentially) only depends on $y$. (b) Mean and spread of the discrete Gaussian curvature on thin ``horizontal" strips. 
}
\label{fig:BranchPointsGaussian}
\end{center}
\end{figure}

Instead of using eqs.~\eqref{eq:Backlund}-\eqref{eq:along-characteristics}, we directly generate discrete {\em Asymptotic nets} \cite{bobenko1999discretization} (allowing for 
potential branch points) that solve (a discretization of) eq.~\eqref{eq:mng_amp} 
-- it is easier to impose boundary conditions in the latter approach.  Figure~\ref{fig:BranchPointsGaussian}(a) illustrates a ``discrete 
isometric immersion'' obtained by introducing branch points into the product solution of eq.~\eqref{eq:mng_amp}. As the step size of the discretization goes to zero, the asymptotic net converges, by construction,  to a surface with a continuously varying tangent plane, {\em i.e.} no ridges or cone-points.  As expected from 
eqs.~\eqref{eq:Backlund}-\eqref{eq:along-characteristics}, our numerical 
algorithm generates a quad-mesh with the same topology as the meshes in fig.~\ref{fig:SelfSimilar}(b). Figure~\ref{fig:BranchPointsGaussian}(b) demonstrates the convergence of the asymptotic net to an isometry of the metric in~\eqref{eq:StripMetric} with $\alpha = 1$ by comparing
the (nondimensional) ``target curvature" $K(y) = -f''(y)$ 
with the piecewise constant discrete Gaussian 
curvature $K_d$ of the asymptotic net.

\begin{figure}[h!]
\begin{center}
\includegraphics[width=\hsize]{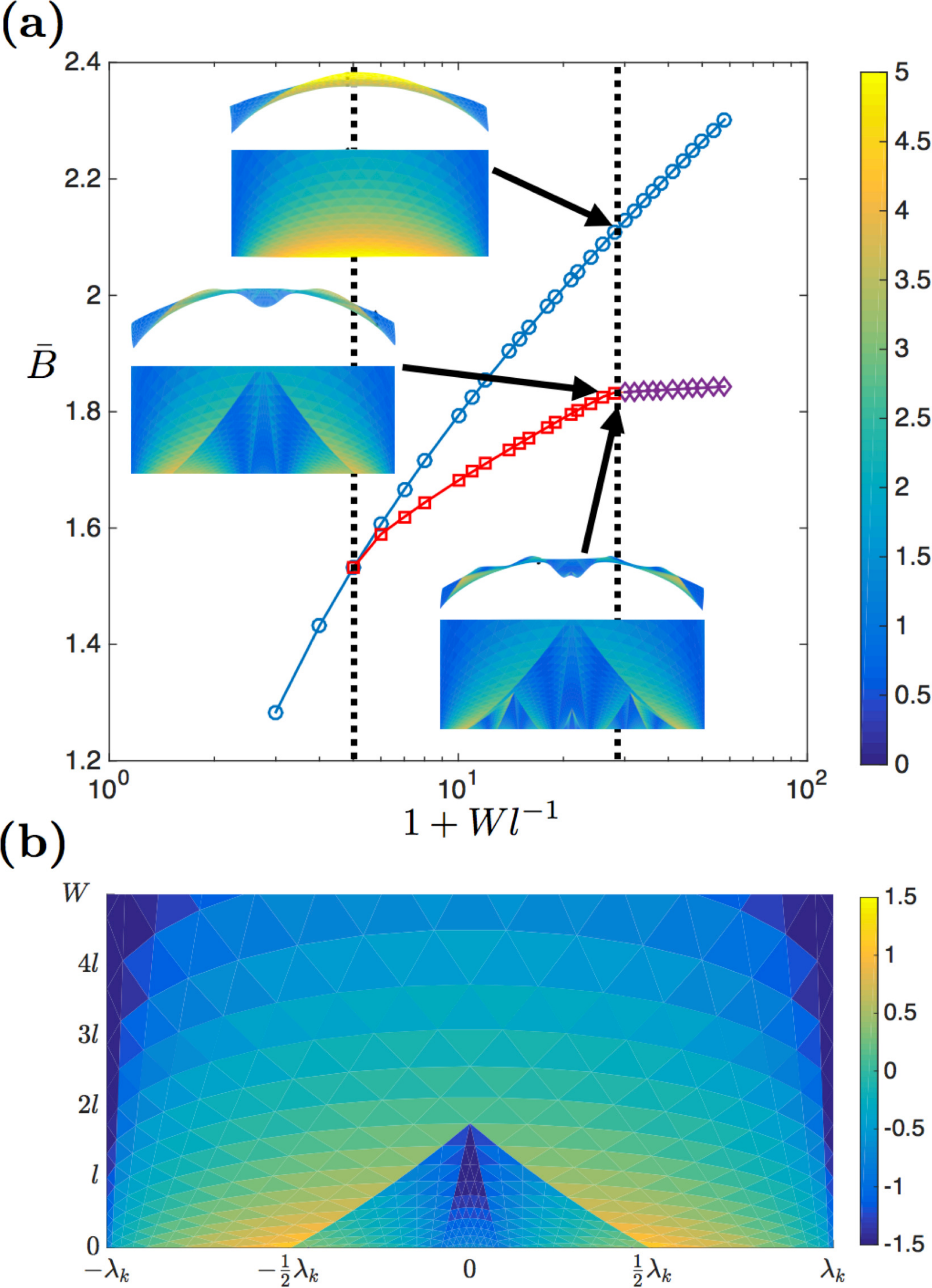}
\caption{(a) Bending energy minimized over a 2-parameter family of isometries for $f(y) = (1+y/l)^{-1}$. Circles (blue)  correspond to single wavelength isometries,  squares (red) to surfaces with one branch point,  diamonds (magenta) to surfaces with four branch points. The insets 
are colored by the normalized energy density $\epsilon^{-2}l^2(4H^2-2K)$. (b) The disparity $\eta$ for the energy minimizing solution with $1+Wl^{-1} = 6$. This figure is to scale. The optimal values are $s_1\approx 9.1, s_2 \approx 4.9, \lambda_k \approx 5.1l$.}
\label{fig:BranchPointsEnergy}
\end{center}
\end{figure}

Given a solution $w(x,y)$ of eq.~\eqref{eq:mng_amp} we obtain a 2-parameter family of rescaled solutions by
$\tilde{w}(x,y) = s_1^{-2} s_2^{\alpha} w(s_1 x, s_2 (y +l)-l)$, with scale factors $s_1 > 0, s_2 \geq 1$.  Figure \ref{fig:BranchPointsEnergy}(a) shows the minimum energy among isometries obtained by rescaling the solution shown in fig.~\ref{fig:BranchPointsGaussian}. As 
$W/l$ increases, 
it is indeed energetically favorable to introduce branch points in $0 \leq y \leq W$; fig.~\ref{fig:BranchPointsEnergy}(b) illustrates that {the first branch point} appears when 
$\eta \simeq 1$ at $(0,0)$. The insets 
in fig.~\ref{fig:BranchPointsEnergy}(a) illustrate the divergence of $H$ with increasing $W/l$ for single wavelength isometries; they also show that introducing branch points does lower  the bending energy.

\section{Exponential metrics and singular edges}
As a special case we now consider the metric:
\begin{equation}\label{eq:ExpMetric}
\mathbf{g}=(1+\epsilon^2 \exp(-2y/l))dx^2+dy^2
\end{equation}
on the {``positive"} half-plane $\mathbb{R}^2_+ = \mathbb{R} \times [0,\infty)$, and also on all of $\mathbb{R}^2$. Equation~\eqref{eq:mng_amp} has a one parameter family of solutions 
$
w^0(x,y)=\sqrt{2}k^{-1}\cos(kx)\exp(-y/l).
$ 
For $k=l^{-1}$, $w^0(x,y)$ is a harmonic function on $\mathbb{R}^2$ and  $H=0$ (a minimal surface) in the small-slope approximation. Consequently, there is no mechanism in the small-slope regime to 
drive sub-wrinkling. The metric~\eqref{eq:ExpMetric} thus provides a natural setting to
 isolate/study nonlinear effects in exact (i.e non small-slope) isometries on wrinkling patterns.

\begin{figure}[ht]
\begin{center}
\includegraphics[width=\hsize]{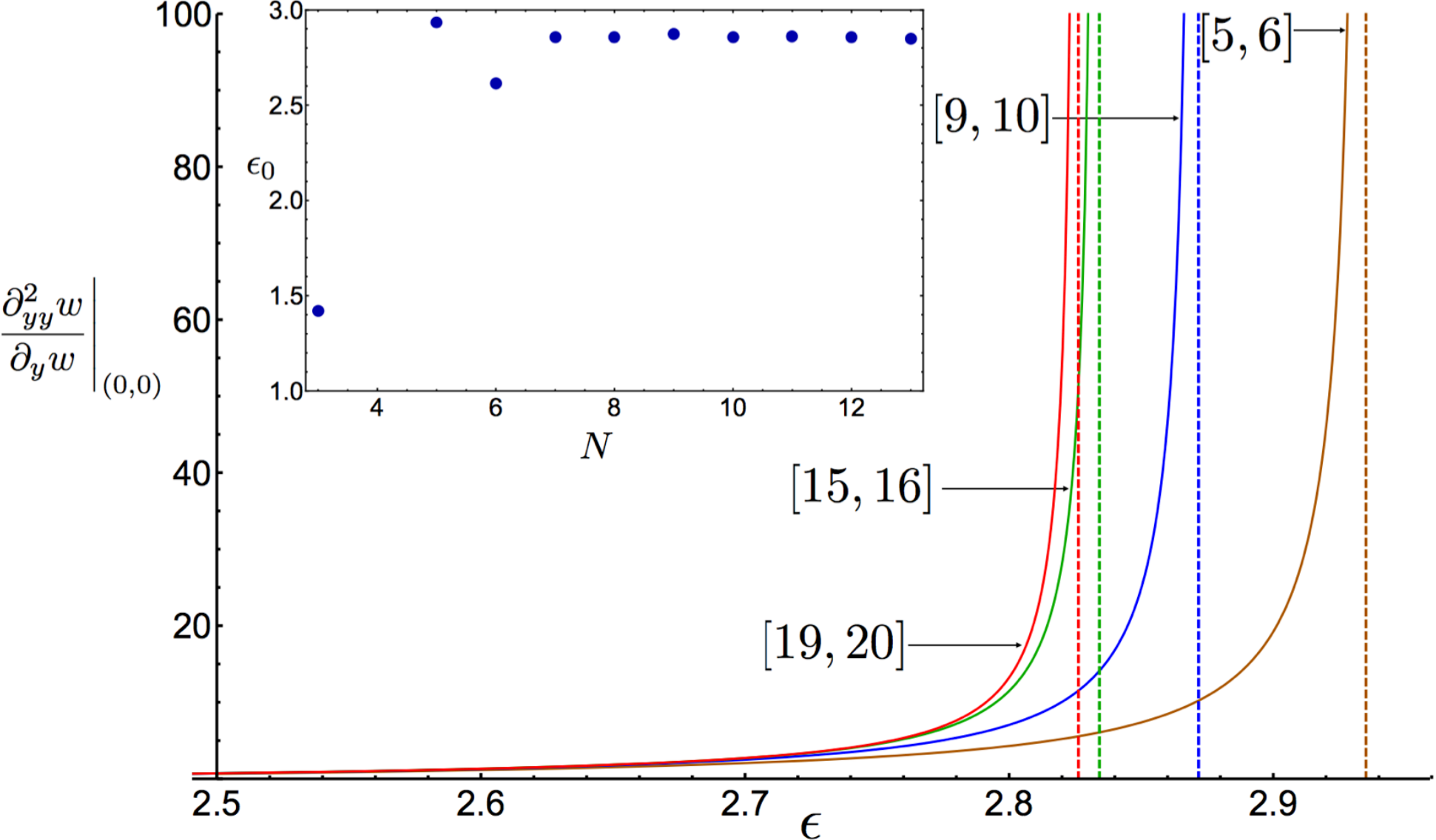}
\end{center}
\caption{$[N,N+1]$ Pad\'{e} approximants to the logarithmic derivative $\partial^2_{yy}w/\partial_yw$  
at $(0,0)$. The inset 
shows the value(s) of $\epsilon_0$ inferred from the vertical asymptote. 
}
\label{fig:DlogPade}
\end{figure}

We can find analytic isometries of the metric by power series expansions: $(x,y) \mapsto (x+\epsilon^2 u,y+ \epsilon^2 v, \epsilon w)$ where  $u=\sum_{i=0}\epsilon^{2i}u^i$, $v=\sum_{i=0} \epsilon^{2i}v^i$  and $w=\sum_{i=0}\epsilon^{2i}w^i$, $w^0$ is as above, {and requiring that the configurations have zero strain at each order in $\epsilon$. 
To extract useful information we need to {\em resum} the series; to this end we use methods based on Pad\'{e} approximants \cite{baker1961application}.

The series solution does not converge on all of $\mathbb{R}^2$. Rather it converges on a region of the form $y > \Pi(x,\epsilon)$. The curve $y = \Pi(x,\epsilon)$ represents the {\em singular edge} \cite{amsler1955surfaces}, which is a {\em horizon} or limit to the extension of an isometry that agrees with $w^0(x,y)$ as $y \to \infty$.}
With increasing $\epsilon$, the singular edge will first intersect with the half-plane $\mathbb{R}^2_+$ at $(0,0)$, {\em i.e.} the maximum for $\left|\partial^2_{yy}w\right|$. 
If $\partial^2_{yy}w(0,y)\sim A(y+\epsilon-\epsilon_0)^{-(\beta+1)}$, we can approximate $\epsilon_0$ and $\beta$ by the poles and residues of the Pad\'{e} approximants to the logarithmic derivative $\partial^2_{yy}w(0,0)/\partial_yw(0,0)$ \cite{baker1961application}. In fig.~\ref{fig:DlogPade} we plot the Dlog Pad\'{e} approximants for the case $kl=1$ and find that $\epsilon_0\approx 2.86$ and $\beta=.500158$. This strongly indicates that a curvature diverges as $s^{-\frac{1}{2}}$, where $s$ is the distance to the singular edge. This divergence is consistent with the singular edge of the pseudosphere and other hyperbolic surfaces of revolution.

\begin{figure}[ht]
\begin{center}
\includegraphics[width=\hsize]{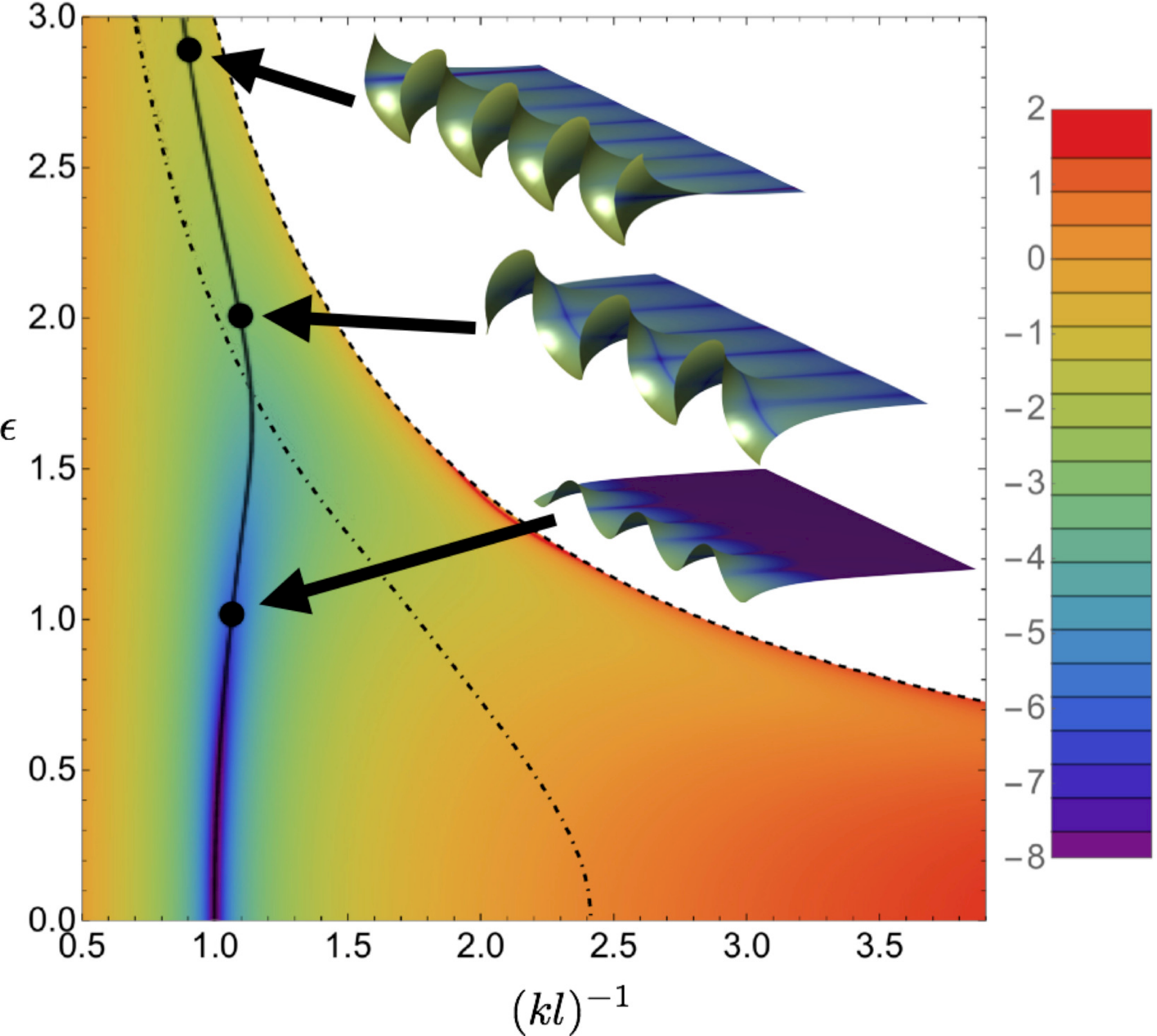}
\end{center}
\caption{Contour plot of  $\ln(\overline{\mathcal{B}})$ as a function of the dimensionless wavelength $(kl)^{-1}$ and $\epsilon$. $\eta = 1$ on the dash-dotted line and the solid line is the energy minimmizing wavelength. The dashed curve indicates values at which the singular curve touches $(0,0)$; it represents a ``horizon".}
\label{fig:BendingEnergy}
\end{figure}

Figure~\ref{fig:BendingEnergy} shows a contour plot of $\overline{\mathcal{B}}[\Delta w]= (2 \pi l)^{-1}k \int_{0}^{2 \pi/k}\int_0^{\infty}( \Delta w)^2\,dy\,dx$ computed from the $[9,9]$ Pad\'{e} approximant of $\Delta w$. The solid line indicates the $\epsilon$ dependent global wavelength $\lambda_{glob} \sim k^{-1}$ that minimizes $\overline{\mathcal{B}}$, a proxy for the bending energy per unit length. For $\epsilon \ll 1$ the full elastic energy selects the expected wavelength from the small slope theory, {\em i.e.} $k=l^{-1}$, while for moderate values of $\epsilon \gtrsim 2$ the singular edge has a dramatic influence on the bending energy. 

The sheet can ``move" the singular edge by a global decrease of  the wavelength, i.e making $k$ larger. While this lowers the bending content near $y=0$, it will introduce large bending content in the rest of the sheet. As in the small-slope case, we expect that the minimizers of the elastic energy will instead locally refine the wavelength by introducing branch points near the edge $y=0$.  The first branch point is expected to occur on the dash-dotted line in fig.~\ref{fig:BendingEnergy} corresponding to the contour $H/\sqrt{|K|}=\eta =1$. The sub-wrinkling threshold for energy minimizers is  at the intersection of the curves, {\em i.e.}, $\epsilon \approx 1.75$. Accounting for in-plane displacement, this corresponds to a slope $\gamma = |\partial_yw|/\sqrt{1-(\partial_yw)^2} \approx 0.67$. Complex buckling patterns can thus occur even in situations where the small-slope theory would predict no sub-wrinkling. Nonlinear contributions to in-plane strains are relevant for slopes as small as $\tan^{-1}(\gamma) \approx 35^\circ$; this is consistent with the ``real world" examples in fig.~\ref{fig:Examples}.

\section{Discussion}
The existence of small-slope isometric immersions for hyperbolic free sheets with finite bending content ensures that the elastic energy per unit thickness  scales as $t^2$. {In the $t \to 0$ limit, this energy is} much smaller than $t^{5/3}$, the energy scale for crumpled sheets \cite{science.paper,lobkovsky}, {and $t^{4/3}$, the energy scale for sub-wrinkling non-Euclidean sheets subject to boundary conditions away from the wrinkled edge \cite{bella2014metric}. Free sheets therefore do not admit elastic ridges \cite{lobkovsky}, $d$-cones \cite{MbAYP97} or wrinkles that balance stretching and bending energies \cite{sharon2007geometrically,bella2014metric} because these defects cost too much energy;} rather, the relevant singularities are {\em branch points} and {\em lines of inflection}. These defects are unique in that they {\em do not concentrate elastic energy} in the vanishing thickness limit. They arise for geometric reasons, {\em viz.} to bypass the {geometric rigidity} of smooth isometric immersions that prevents the refinement of the pattern wavelength. To our knowledge, this is the first example of a condensed matter system that is driven by {\em geometric} rather than {\em energetic} defects.

We have shown that  the buckling patterns in free sheets are a result of competition between the contributions to the bending energy from the two principal curvatures. Our key contribution is in identifying  a mechanism,  {\em viz.} the introduction of branch points/lines of inflection, that lowers the bending energy of the sheet while preserving the isometry constraint and the continuity of the tangent plane. 

{Audoly and Boudaoud~\cite{audoly2003self}  suggested the possibility that sub-wrinkling can arise in isometries. Their predicted scaling of the wavelength with the thickness implies that the limiting isometries have infinite bending content.  The mechanism elucidated in this paper, {\em viz.} sub-wrinkling through introduction of branch points, is of an entirely different character, and outside the realm of the smooth truncated Fourier series solutions considered in~\cite{audoly2003self}.} 

We {have analyzed our proposed mechanism} and give a quantitative criterion for incipient sub-wrinkling in free sheets. 
{The far from threshold behavior is determined by a global optimization of the bending energy. We do not yet have criteria for the number/locations of multiple branch points in this regime. Our general, heuristic, conclusions are -- (i) wide ($W \gtrsim l$) hyperbolic elastic sheets with an $O(1)$ variation in the logarithm of the Gaussian curvature $\log|K(y)|$ will generically manifest branch points and lines of inflection; fractal-like, sub-wrinkling, profiles have lower energies than their single wavelength counterparts, and  (ii) nonlinear contributions to in-plane strain, which one might otherwise consider small, also promote sub-wrinkling; this effect is already evident for slopes $\lesssim 1$.}
 
 This work is primarily concerned with the vanishing thickness limit $t \to 0$. The results however are immediately transferable to physical sheets.  In a physical sheet, the lines of inflection and branch points are  replaced by boundary layers; the sheet configuration is indeed smooth, albeit with small scale structures. We expect that the  smoothed defects are akin to the $t^{1/3}$ and $t^{1/2}$ boundary layers that resolve the non-smoothness of the $n$-saddle isometries of disks with $K=-1$ \cite{gemmer2012defects}.
 
Our construction  shows  there are {\em continuous families} of low-energy states obtained by appropriately 
gluing together isometries. Variations within these families lead to  ``floppy modes" of deformation. Thin hyperbolic free sheets are thus easily deformed by weak stresses. The buckling pattern may be sensitive to the dynamics of the swelling process, experimental imperfections, or other external forces. A statistical description of the singularities and their interactions is therefore a natural approach to study this system. 

\acknowledgments
{We are very grateful to the referees for their input.} JG, ES and SV were supported by an US-Israel BSF grant 2008432. JG and SV were also supported  by the NSF grant DMS-0807501. JG is currently supported by NSF-RTG grant DMS-1148284. This research was supported in part by the NSF through Grant No. NSF PHY11-25915. 

\bibliographystyle{eplbib}
\bibliography{ref}

\end{document}